\documentclass[aps,pra,reprint,a4paper]{revtex4-1}
\usepackage{bm}
\usepackage[dvipdfmx]{graphicx}
\usepackage{dcolumn}
\usepackage{color}
\usepackage{amsmath}
\usepackage{amssymb}
\usepackage{ulem}

\begin{document}


\title{The structure of approximate two electron
wavefunctions in intense laser driven ionization dynamics}



\author{Takeshi Sato}
\email[Electronic mail:]{sato@atto.t.u-tokyo.ac.jp}

\author{Kenichi L. Ishikawa}
\email[Electronic mail:]{ishiken@n.t.u-tokyo.ac.jp}
\affiliation{
Photon Science Center, School of Engineering, 
The University of Tokyo, 7-3-1 Hongo, Bunkyo-ku, Tokyo 113-8656, Japan
}
\affiliation{
Department of Nuclear Engineering and Management, School of Engineering,
The University of Tokyo, 7-3-1 Hongo, Bunkyo-ku, Tokyo 113-8656, Japan}



\begin{abstract}
The structure of approximate two electron wavefunction is
deeply investigated, both theoretically and numerically, in the
strong-field driven ionization dynamics.
Theoretical analyses clarify that for two electron singlet systems, the previously proposed
time-dependent extended Hartree-Fock (TD-EHF) method 
[Phys. Rev. A {\bf 51}, 3999 (1995)]
is equivalent to the multiconfiguration time-dependent Hartree-Fock method with two
occupied orbitals.
The latter wavefunction is further transformed into the
natural expansion form, enabling the direct propagation of the natural
orbitals (NOs).
These methods, as well as the conventional time-dependent Hartree-Fock
(TDHF) method, are numerically assessed for the description of ionization 
dynamics of one-dimensional helium atom model. This numerical analysis
(i) explains the reason behind the well-known failure of TDHF method 
to describe tunneling ionization, 
(ii) demonstrates the interpretive power of the TD-EHF wavefunction 
both in the original nonorthogonal and the NO-based formulations, and 
(iii) highlights different manifestations of the electron correlation (effect
beyond the single determinant description), in tunneling ionization,
high harmonic generation, and nonsequential double ionization.
Possible extensions of the NO basis approach to multielectron systems
are briefly discussed.

\end{abstract} 


\maketitle 

\section{Introduction\label{sec:introduction}}
Tunneling ionization (TI) is one of the most important processes in the fields
of high-field and ultrafast physics.
It is a purely quantum mechanical effect, where electrons in an
atom or molecule escape the binding potential with a finite
probability by tunneling through the potential barrier distorted by an intense
electromagnetic field. In case of an oscillating field, 
the ejected electron, feeling a reverse acceleration, may return to the
core region, possibly undergoing recombination with the parent ion, or
recollision with it to induce further excitation or ionization. 
These electron dynamics in the ultrashort, intense, and oscillating fields
induce a variety of important nonlinear, nonperturbative phenomena, such as
high harmonic generation (HHG) and nonsequential double ionization (NSDI).

Although the time-dependent Schr\"odinger equation (TDSE) provides the rigorous
theoretical framework [within a single atom (molecule) response]
for investigating such electron dynamics
\cite{Pindzola1998PRA,Pindzola1998JPB,Colgan2001JPB, 
Parker2001,Laulan2003PRA,Piraux2003EPJD,Laulan2004PRA,ATDI2005,
Feist2009PRL,Pazourek2011PRA,He_TPI2012PRL,Suren2012PRA,He_TPI2013AS,
Vanroose2006PRA,Horner2008PRL,Lee2010JPB}, direct applications of TDSE
for systems with more than two electrons are extremely difficult.
Therefore, the single-active electron (SAE) approximation has been
widely used, in which only the outermost electron is explicitly treated,
with the effect of all the other electrons embedded in a fixed model potential.
This approximation, however, fails to account for multielectron effects
in high-field phenomena. 
As a result, the correlated electron dynamics in, e.g., NSDI process
is outside the scope of this approximation.
Moreover, for the high-harmonic spectroscopy of multielectron
dynamics in molecules \cite{Smirnova:2009, Smirnova2009PNAS,
Mairesse:2010}, the effect of multichannel ionization has to be taken
into account beyond the SAE picture.

A number of time-dependent wavefunction methods have been proposed to
describe the multielectron dynamics in intense laser fields.
The time-dependent Hartree-Fock (TDHF) method, which approximates the
total wavefunction with a single determinant, was first applied to the  
ionization dynamics by Kulander \cite{Kulander:1987}.
Unfortunately as revealed by Pindzola {\it et al} \cite{Pindzola:1991},
the TDHF method gives a qualitatively wrong
description of the ionization process, despite the fact that the (time
independent) HF method generally describes
field-free ground-state of atoms and molecules quite well.
A breakthrough was the introduction of the multiconfiguration
time-dependent Hartree-Fock (MCTDHF) method \cite{Caillat2005PRA, Kato:2004},
which extends the wavefunction {\it ansatz} to the linear superposition of
determinants with both expansion coefficients and constituent one
electron wavefunctions (orbitals) treated as variational degrees of
freedom. We recently proposed a more flexible time-dependent theory \cite{Sato:2013}
based on the concept of complete-active-space self-consistent-field (CASSCF) 
originally developed in quantum chemistry \cite{Roos:1987}, to extend
the applicability of the MCTDHF method. See Ref.~\cite{Sato:2013} for
more comprehensive review of the proposed theoretical approaches
including very recent developments.

With increase in expansion length, the
MCTDHF method provides a powerful series of
approximations from the single determinant TDHF to the exact
TDSE limit.
However, one should recognize that, apart from the severe exponential
scaling of the computational cost against the system size, 
the interpretation of numerical results gets more and more difficult for
simulations with higher accuracy, where the wavefunction is expanded with
numerous determinants.
It is, therefore, important to establish as simple an approximation as
possible which captures essential features of electron dynamics
while keeping the conceptual simplicity to allow for a 
clear interpretation of simulation results and deep insight into the
physics involved.

In this work, we present theoretical and numerical investigations on the
structure of approximate two electron wavefunction---the simplest multielectron
system---and their capability to describe intense laser driven ionization
dynamics. We discuss the simplest form of the wavefunction required for a
physically correct description of the TI process.
Specifically, we focus on the time-dependent extended Hartree-Fock
(TD-EHF) method \cite{Pindzola:1995, Pindzola:1997, Tolley:1999,
Dahlen:2001, Nguyen:2006}, which approximates the total
wavefunction by a symmetrized product of nonorthogonal orbitals [Eq.~(\ref{eq:ehf})]. 
We establish a transformation of the EHF wavefunction into a more
convenient expression in terms of the orthonormal natural orbitals
(NOs). This transformation enables the systematic improvement of
approximations on top of the TD-EHF method, and also opens the
possibility of its generalization to multielectron (beyond two electron) systems.

This paper is organized as follows. In Sec.~\ref{sec:theory} we
theoretically investigate the TDHF and TD-EHF methods both in the
original nonorthogonal as well as the NO-based formulations, and derive the
equations of motion (EOMs) for the latter.
Then in Sec.~\ref{sec:numerical}, we apply various theoretical methods to the
dynamics of one-dimensional helium model to investigate their
performance in describing TI, HHG and NSDI processes. 
Finally, the summary of this work and some future prospects are given 
in Sec.~\ref{sec:summary}.
The Hartree atomic units are used throughout unless otherwise noted.

\section{Theoretical analyses of the two electron
 wavefunction\label{sec:theory}}

\subsection{RHF, UHF, and EHF wavefunctions\label{subsec:rhf_uhf_ehf}}

As mentioned in Sec.~\ref{sec:introduction}, the TDHF method based on the
the restricted HF (RHF) ansatz cannot describe the TI process
correctly. 
For singlet two electron systems, the RHF wavefunction $\Psi_{\rm RHF}$
and the EOM for the orbital (TD-RHF) are given by,
\begin{eqnarray}\label{eq:rhf}
\Psi_\mathrm{RHF} =
\Vert\psi_1(\bm{r}_1,t)\bar{\psi}_1(\bm{r}_2,t)\Vert,
\end{eqnarray}
\begin{eqnarray}\label{eq:td-rhf}
\mathrm{i}\dot{\psi}_1(\bm{r},t) \! = \!
\left[
h(\bm{r},t)  + J[{\psi_1}](\bm{r})
\right] \psi_1(\bm{r},t),
\end{eqnarray}
respectively, where the symbol $\Vert \cdot\cdot\cdot \Vert$ denotes a normalized
Slater determinant, with $\psi_1$ ($\bar{\psi}_1$) being 
the direct product of a spatial orbital function
$\psi(\bm{r})$ and an up (down) spin eigenfunction.
The operator $h$ is the one-electron part of the total
Hamiltonian, and $J$ is the Coulomb operator defined in
Eq.~(\ref{eq:coulomb}).
The TD-RHF method cannot describe the spatially different
motions of the ionizing electron and that left in the ionic core, since
it enforces the closed-shell structure.

One may consider that the single determinant unrestricted Hartree-Fock (UHF)
wavefunction,
\begin{eqnarray}\label{eq:uhf}
\Psi_\mathrm{UHF} = \Vert\psi_1(\bm{r}_1,t)\bar{\psi}_2(\bm{r}_2,t))\Vert,
\end{eqnarray}
fits the purpose of describing the different motions of two electrons.
However, if the initial two orbitals coincide,
$\psi_1(\bm{r},0) = \psi_2(\bm{r},0)$, as is often the case for stable ground states with
even number of electrons \cite{Szabo:1996}, these orbitals remain the
same at any later time, $\psi_1(\bm{r},t) = \psi_2(\bm{r},t)$, unless exposed to a
spin-dependent perturbation. This is understood by seeing that the TD-UHF equations,
\begin{subequations}\label{eq:td-uhf}
\begin{eqnarray}
\mathrm{i}\dot{\psi}_1(\bm{r},t) &=& \left\{
h(\bm{r},t) + J[{\psi_2}]\right\} \psi_1(\bm{r},t), \\
\mathrm{i}\dot{\psi}_2(\bm{r},t) &=& \left\{
h(\bm{r},t) + J[{\psi_1}]\right\} \psi_2(\bm{r},t),
\end{eqnarray}
\end{subequations}
are symmetric for two spatial functions.
Thus the TD-UHF approach cannot solve the problem of the
TD-RHF method. More seriously, the
UHF wavefunction is generally not the spin eigenfunction.
The expectation value of the total spin operator is given by
(assuming normalized orbitals)
\begin{eqnarray}\label{eq:s2_uhf}
\langle \Psi_{\rm UHF} \vert S^2 \vert \Psi_{\rm UHF} \rangle
= 1 - \vert \langle \psi_1 \vert \psi_2 \rangle \vert^2.
\end{eqnarray}
If two orbitals are orthogonal, as expected, e.g., in the
homogeneous dissociation of a hydrogen molecule or in the single
ionization limit of helium, the UHF wavefunction corresponds to 
an equal weight mixture of singlet and triplet. The dynamics should be
strongly altered by this spin contamination.

Clearly, the simplest allowed wavefunction capable of describing
ionization processes consists of the 
symmetrized product of two different spatial orbitals, 
\begin{eqnarray}\label{eq:ehf}
\Psi_{\rm EHF} 
= \frac{1}{\sqrt{2}} \left\{
\psi_1(\bm{r}_1,t)\psi_2(\bm{r}_2,t) + 
\psi_2(\bm{r}_1,t)\psi_1(\bm{r}_2,t)
\right\}, \nonumber \\
\end{eqnarray}
multiplied by a singlet spin function.
Here $\psi_1$ and $\psi_2$ has to be nonorthogonal to seamlessly
describe the closed-shell dominant ground state and open-shell dominant
excited and continuum states.
This form of wavefunction was used, in a phenomenological
formulation, to explain the mechanism of NSDI \cite{Watson:1997}.
In more rigorous variational treatments \cite{Pindzola:1995,
Pindzola:1997, Tolley:1999, Dahlen:2001, Nguyen:2006}, 
this is often called the extended Hartree-Fock (EHF) wavefunction, and
has been applied to electron dynamics of two electron systems.
Although the same wavefunction has been sometimes called UHF
wavefunction \cite{Pindzola:1995, Pindzola:1997, Tolley:1999}, we adopt
the term EHF in this work to avoid confusion with the definition of the
UHF in Eq.~(\ref{eq:uhf}). 

\subsection{TD-EHF method\label{subsec:td-ehf}}

Following Tolley \cite{Tolley:1999}, but adopting concise notations of
Nguyen and Bandrauk \cite{Nguyen:2006}, the TD-EHF equation is written
in the matrix form as follows; 
\begin{eqnarray}\label{eq:td-ehf}
{\rm i} \dot{\bm{u}} = 
\left\{(h - W)\bm{1} + \bm{S}^{-1} \bm{V} \right\} \bm{u},
\end{eqnarray}
where $\bm{u}$ is a column vector function with elements 
$\bm{u}(\bm{r}) = \left(\psi_2(\bm{r}),\psi_1(\bm{r})\right)^{\rm T}$,
$\bm{1}$ is a two by two unit matrix, and 
\begin{eqnarray}\label{eq:ehf_ovlp}
\bm{S} = \left(
\begin{array}{ll}
\langle\psi_1\vert\psi_1\rangle & \langle\psi_1\vert\psi_2\rangle \\
\langle\psi_2\vert\psi_1\rangle & \langle\psi_2\vert\psi_2\rangle \\
\end{array}
\right) = \left(
\begin{array}{ll}
1         & \lambda \\
\lambda^* & 1       \\
\end{array}
\right),
\end{eqnarray}
\begin{eqnarray}\label{eq:ehf_meanfield}
\bm{V} = \left(
\begin{array}{cc}
J[{\psi_1}] + K[{\psi_1}] & -\alpha_{12} \\
-\alpha_{21} & J[{\psi_2}] + K[{\psi_2}] \\
\end{array}
\right),
\end{eqnarray}
\begin{eqnarray}\label{eq:ehf_psene}
W = 
\frac{
\langle \psi_1\psi_2 \vert\vert \psi_1\psi_2 \rangle -
\lambda  \langle \psi_2\psi_2 \vert \psi_2\psi_1 \rangle  -
\lambda^* \langle \psi_1\psi_1 \vert \psi_1\psi_2 \rangle
}
{\det(\bm{S})}, \nonumber \\
\end{eqnarray}
where
$\alpha_{12} = \langle \psi_1\psi_1 \vert \psi_1\psi_2 \rangle - W \lambda,
 \alpha_{21} = \langle \psi_2\psi_2 \vert \psi_2\psi_1 \rangle - W
 \lambda^*$, and
$\langle \psi_1\psi_2 \vert\vert \psi_1\psi_2 \rangle \equiv
\langle \psi_1\psi_2 \vert \psi_1\psi_2 \rangle +
\langle \psi_1\psi_2 \vert \psi_2\psi_1 \rangle$,
with a shorthand notation
for two electron repulsion integrals:
\begin{eqnarray}\label{eq:int2e}
\langle \chi_1\chi_2 \vert \chi_3\chi_4 \rangle =
\int \!\! d\bm{r}_1 d\bm{r}_2 \frac{
\chi^*_1(\bm{r}_1)\chi^*_2(\bm{r}_2)
\chi_3(\bm{r}_1)\chi_4(\bm{r}_2)}{\vert \bm{r}_1 - \bm{r}_2 \vert},
\nonumber \\
\end{eqnarray}
for a given orbital quartet $\chi_1,\chi_2,\chi_3,\chi_4$.
The Coulomb $J$ and exchange $K$ operators appearing in
Eq.~(\ref{eq:ehf_meanfield}) are defined by
\begin{eqnarray}\label{eq:coulomb}
J[{\chi}](\bm{r})
= \int\!\!d\bm{\bar{r}} 
\frac{\chi^*(\bm{\bar{r}}) \chi(\bm{\bar{r}})}
{|\bm{r} - \bm{\bar{r}}|},
\end{eqnarray}
\begin{eqnarray}\label{eq:exchange}
\left(K[{\chi}]\psi\right)(\bm{r})
= \int\!\!d\bm{\bar{r}} 
\frac{\chi^*(\bm{\bar{r}}) \psi(\bm{\bar{r}})}
{|\bm{r} - \bm{\bar{r}}|} \chi(\bm{r}),
\end{eqnarray}
for given orbitals ${\chi}$ and $\psi$.
We have adopted the convention \cite{Pindzola:1995, Pindzola:1997, Tolley:1999} that
$\psi_1$ and $\psi_2$ are normalized, and nonorthogonal to each other with overlap
$\langle\psi_1\vert\psi_2\rangle = \lambda$. Thus, the total wavefunction is not
normalized but $\langle\Psi_{\rm EHF}\vert\Psi_{\rm EHF}\rangle = 1 + \vert
\lambda \vert^2$. This is mathematically equivalent to an apparently different
approach in Refs.~\cite{Dahlen:2001, Nguyen:2006} where the orbitals are
not normalized but the total wavefunction is normalized. 

\subsection{EHF wavefunction in terms of orthogonal orbitals
\label{subsec:ehf_orth}}
A problem of the TD-EHF method is the difficulty to improve the accuracy
beyond a {\it single} antisymmetrized product in Eq.~(\ref{eq:ehf}), due
to the use of nonorthogonal orbitals. Thus it is desirable to formulate an
equivalent theory in terms of orthogonal orbitals.
To show that this is in fact possible, we apply the canonical orthogonalization to the
nonorthogonal orbitals $\bm{\psi} = (\psi_1,\psi_2)$ to obtain an
orthonormal set $\bm{\phi} = (\phi_1,\phi_2)$,
\begin{eqnarray}\label{eq:trans}
\bm{\phi}(\bm{r},t) = \bm{\psi}(\bm{r},t) \bm{X}(t), \hspace{.5em}
\bm{X} = \bm{U} \bm{s}^{-1/2},
\end{eqnarray}
where the unitary matrix $\bm{U}$ 
diagonalizes the overlap matrix $\bm{S}$ of Eq.~(\ref{eq:ehf_ovlp}) and
$\bm{s}$ is the diagonal matrix with elements being corresponding eigenvalues
$s_{\pm} = 1 \pm \vert\lambda\vert$.
Upon this transformation, the EHF wavefunction of Eq.~(\ref{eq:ehf}) is
expressed in terms of orthogonal orbitals $\{\phi_1,\phi_2\}$ as
\begin{eqnarray}\label{eq:ehf_no}
\Psi_{\rm EHF} &=&
A_1(t) \Vert\phi_1(\bm{r}_1,t) \bar{\phi}_1(\bm{r}_2,t)\Vert \nonumber \\ &-&
A_2(t) \Vert\phi_2(\bm{r}_1,t) \bar{\phi}_2(\bm{r}_2,t)\Vert,
\end{eqnarray}
where 
\begin{subequations}
\begin{eqnarray}\label{eq:ehf_transf_ci}
A_1 &=& \frac{1 + \vert\lambda\vert}{\sqrt{2(1 + \vert\lambda\vert^2)}}
         \frac{\lambda^*}{\vert\lambda\vert}, \\
A_2 &=& \frac{1 - \vert\lambda\vert}{\sqrt{2(1 + \vert\lambda\vert^2)}}
         \frac{\lambda}{\vert\lambda\vert},
\end{eqnarray}
\end{subequations}
\begin{subequations}
\begin{eqnarray}\label{eq:ehf_transf_mo}
\phi_1(\bm{r}) &=& \frac{1}{\sqrt{2(1 + \vert\lambda\vert)}}
\left\{\frac{\lambda}{\vert\lambda\vert} \psi_1(\bm{r}) + \psi_2(\bm{r})\right\}, \\
\phi_2(\bm{r}) &=& \frac{1}{\sqrt{2(1 - \vert\lambda\vert)}}
\left\{\psi_1(\bm{r}) - \frac{\lambda^*}{\vert\lambda\vert} \psi_2(\bm{r}) \right\}.
\end{eqnarray}
\end{subequations}

Inversely, provided that the total wavefunction is given in the form of
Eq.~(\ref{eq:ehf_no}), it can be transformed back into
Eq.~(\ref{eq:ehf}) through $\bm{\psi}(\bm{r},t) = \bm{\phi}(\bm{r},t)
\bm{X}^{-1}(t)$ as (assuming $\vert A_1 \vert > \vert A_2 \vert$)
\begin{subequations}\label{eq:ehf_transb}
\begin{eqnarray}
\label{eq:ehf_transb_psi1}
\psi_1(\bm{r}) &=&
\sqrt{\frac{1 + \vert\lambda\vert}{2}} \frac{\lambda^*}{\vert\lambda\vert} \phi_1(\bm{r}) +
\sqrt{\frac{1 - \vert\lambda\vert}{2}} \phi_2(\bm{r}),
\\
\label{eq:ehf_transb_psi2}
\psi_2(\bm{r}) &=&
\sqrt{\frac{1 + \vert\lambda\vert}{2}} \phi_1(\bm{r}) -
\sqrt{\frac{1 - \vert\lambda\vert}{2}} \frac{\lambda}{\vert\lambda\vert} \phi_2(\bm{r}),
\\
\label{eq:ehf_transb_ovlp}
\lambda &=& 
\frac{\vert A_1\vert - \vert A_2\vert}{\vert A_1\vert + \vert A_2\vert}
\left(\frac{\vert A_2/A_1 \vert}{A_2/A_1}
\right)^{1/2}.
\end{eqnarray}
\end{subequations}
The existence of the reversible map $\bm{X}$ between the two
expressions of the total wavefunctions, Eqs.~(\ref{eq:ehf}) and
(\ref{eq:ehf_no}), demonstrates that the electron dynamics can be
equivalently represented with either expression. Thus instead of
solving Eq.~(\ref{eq:td-ehf}), we can formulate an equivalent
theory by applying time-dependent variational method to
Eq.~(\ref{eq:ehf_no}) with both orbitals $\{\phi_1(t),\phi_2(t)\}$ and coefficients
$\{A_1(t), A_2(t)\}$ treated as variational degrees of freedom. This
will be done in Sec.~\ref{subsec:td-no} after discussing the natural
expansion of two electron wavefunctions.

\subsection{Natural expansion of two electron wavefunction
\label{subsec:natural_2e}}

The most general expansion of two electron wavefunction with given
number, $n$, of orthonormal spatial orbitals is
\begin{eqnarray}\label{eq:mctdhf}
\Psi_{{\rm MC}} = \sum_{i=1}^n\sum_{j=1}^n C_{ij}(t) 
\Vert \phi^\prime_i(\bm{r}_1,t) \bar{\phi}^\prime_j(\bm{r}_2,t) \Vert.
\end{eqnarray}
This corresponds to the wavefunction used in the MCTDHF method.
Now we show that this wavefunction can be reduced to the diagonal form:
\begin{eqnarray}\label{eq:mctdhf_no}
\Psi_{{\rm NO}} = \sum_{i=1}^n A_i(t) 
\Vert \phi_i(\bm{r}_1,t) \bar{\phi}_i(\bm{r}_2,t) \Vert,
\end{eqnarray}
in the case of singlet.
The proof is made by noting that 
the matrix  $\bm{C}$ with elements $C_{ij}$ in Eq.~(\ref{eq:mctdhf})
is complex symmetric, thus can always be factorized to the diagonal form
$\bm{A}$ with diagonal elements $A_i$ in Eq.~(\ref{eq:mctdhf_no}), by
Takagi's factorization \cite{Takagi:1927, Bunse-Gerstner:1988}:
\begin{eqnarray}\label{eq:takagi}
\bm{C} = \bm{V} \bm{A} \bm{V}^{\rm T}, \hspace{.5em}
\bm{A} = \bm{V}^\dagger \bm{C} \bm{V}^*,
\end{eqnarray}
where upper scripts ${\rm T}$, ${\dagger}$, and $*$ stand for
transpose, Hermitian conjugate, and complex conjugate, respectively.
Two sets of orthonormal orbitals in Eqs.~(\ref{eq:mctdhf}) and
(\ref{eq:mctdhf_no}) are connected by the unitary transformation $\bm{V}$
\begin{eqnarray}\label{eq:trans_no}
\phi_i(\bm{r}) = \sum_j \phi^\prime_j(\bm{r}) V_{ji}.
\end{eqnarray}

The orbitals $\{\phi_i\}$ in Eq.~(\ref{eq:mctdhf_no}) have a special
significance of being the natural orbitals (NOs), i.e., 
diagonalize the first order density matrix (1RDM): 
\begin{subequations}\label{eq:eq:1rdm}
\begin{eqnarray}
\rho(\bm{r}_1,\bm{r}^\prime_1) &=& 2\int\!d\bm{r}_2 
\label{eq:1rdm_def}
\Psi(\bm{r}_1,\bm{r}_2) \Psi^*(\bm{r}^\prime_1,\bm{r}_2) \\
\label{eq:1rdm_no}
&=& 2\sum_{i=1}^n \vert A_i \vert^2 \phi_i(\bm{r}_1)\phi^*_i(\bm{r}^\prime_1),
\end{eqnarray}
\end{subequations}
and expansion coefficients are directly connected to
the natural occupation numbers (eigenvalues of 1RDM), 
\begin{eqnarray}\label{eq:occupation} 
\eta_i = 2\vert A_i\vert^2.
\end{eqnarray}
Hereafter we call $\{A_i\}$ the natural coefficients (NCs).
The equivalence of Eqs.~(\ref{eq:ehf}) and (\ref{eq:ehf_no}), as well as
of Eqs.~(\ref{eq:mctdhf}) and (\ref{eq:mctdhf_no}), are known in
quantum chemistry \cite{Szabo:1996,Kutzelnigg:1963}, for real stationary
wavefunctions. Here we have explicitly shown that the same relations
hold for arbitrary (both stationary and non-stationary) wavefunctions. 
In what follows, we will discuss the significance of these transformations
in the time-dependent simulation and interpretation of two electron dynamics.

\subsection{Direct propagation of natural orbitals
\label{subsec:td-no}}
The natural expansion [Eq.~(\ref{eq:mctdhf_no})] is a special (NO)
representation of the general MCTDHF wavefunction [Eq.~(\ref{eq:mctdhf})].
Therefore, the EOMs for the former wavefunction can be derived by
using the invariance of the latter wavefunction
with respect to the unitary transformation among the occupied orbitals.
We first write down the MCTDHF equation [Eqs.~(30) and (40) of
Ref.~\cite{Sato:2013}, e.g.] for the two electron system;
\begin{eqnarray}
\label{eq:mctdhf_ci}
{\rm i} \dot{C}_{ij} &=& \sum_{kl} \left(
h_{ik}\delta_{jl} + h_{jl}\delta_{ik}
+ \langle ij \vert kl \rangle
\right) C_{kl},
\end{eqnarray}
\begin{eqnarray}
\label{eq:mctdhf_orb}
{\rm i}\dot{\phi}_i &=&
\hat{Q} \left(h\phi_i + \sum_j{\Gamma}_j \phi_j \rho^{-1}_{ji}\right)
+ \sum_{j\neq i} \phi_j R_{ji},
\end{eqnarray}
where $h_{ij} = \langle\phi_i\vert h\vert\phi_j\rangle$,
$\langle ij\vert kl\rangle = \langle\phi_i\phi_j\vert\phi_k\phi_l\rangle$,
$\hat{Q} \equiv 1 - \sum_j\vert\phi_j\rangle\langle\phi_j\vert$,
$\Gamma_i \phi_i = 2 \sum_{jkl} W_{jk} \phi_l C^*_{ij}C_{lk}$, and
\begin{eqnarray}\label{eq:meanfield}
W_{ij}(\bm{r}) = \int\!d\bm{\bar{r}} 
\frac{\phi^*_i(\bm{r}) \phi_j(\bm{r})}
{\vert \bm{r} - \bm{\bar{r}} \vert}.
\end{eqnarray}
In Eq.~(\ref{eq:mctdhf_orb}) the matrix $R$ with elements $R_{ij} \equiv
{\rm i}\langle\phi_i\vert\dot{\phi}_j\rangle$ can
be an arbitrary Hermitian matrix in the general MCTDHF formulation
\cite{Sato:2013}. Here this freedom is used to impose a
condition that the 1RDM in the orbital basis $\rho_{ij}$ is
kept diagonal, $d\rho_{ij}/dt = 0\hspace{.5em}(i \neq j)$. Following
Refs.~\cite{Jansen:1993, Manthe:1994} 
which discuss the natural expansion of bosonic wavefunction, we obtain 
the EOMs for NCs and NOs as follows;
\begin{subequations}\label{eq:td-no}
\begin{eqnarray}
\label{eq:td-no_nc}
{\rm i} \dot{A}_i &=& \sum_{j=1}^n \left(
2\delta_{ij} h_{ii} + \langle ii \vert jj \rangle
\right) A_j, \\
\label{eq:td-no_no}
{\rm i}\dot{\phi}_i &=& 
\hat{Q} {F}_i \phi_i \eta^{-1}_i
+ \sum_{j\neq i} \phi_j R_{ji},
\end{eqnarray}
\end{subequations}
where 
\begin{eqnarray}\label{eq:td-no_fock}
F_i \phi_i = h \phi_i \eta_i + 2 \sum_j W_{ij} \phi_j A^*_iA_j,
\end{eqnarray}
and the orbital rotation matrix $R$ is identified as
\begin{eqnarray}
R_{ij} &=& \frac{F_{ij} - F^*_{ji}}{n_j - n_i}, \hspace{.5em}
F_{ij} = \langle\phi_i\vert F_j \phi_j\rangle.
\end{eqnarray}

For notational brevity, we call this approach TD-NO$n$, with $n$ denoting the
number of NOs. The orthogonal reformulation of the EHF wavefunction,
Eq.~(\ref{eq:ehf_no}), is a special case with $n = 2$. An important
advantage of the TD-NO$n$ is its capability of improving the accuracy by
increasing $n$. Moreover it allows the extension to multielectron
systems as discussed below in Sec.~\ref{sec:summary}.
We emphasize that all the methods discussed in this section
are not phenomenological but based on the physically solid variational principle.

\section{Numerical assessments\label{sec:numerical}}
In this section, we numerically investigate the various ansatz of two 
electron wavefunctions discussed in Sec.~\ref{sec:theory}.
For this purpose, we use the one-dimensional helium (1D-He) model.
The electronic Hamiltonian is given by
\begin{eqnarray}\label{eq:ham_1d2e}
H = \sum_{p=1}^2 \left[ -\frac{\partial^2}{\partial z^2_i}
- \frac{2}{\|z_i\|}
- z_i E(t)\right]
+ \frac{1}{\|z_1-z_2\|},
\end{eqnarray}
for two electronic coordinates $z_1$ and $z_2$, where
the nuclear potential is centered at the origin, and soft Coulombic
operator $1/\|x\| \equiv 1/\sqrt{x^2+1}$
is used both for nucleus-electron and electron-electron
interactions.
The electron-laser interaction is included within the dipole
approximation and in the length gauge.
Note that all the methods examined in this work are gauge-invariant as
discussed, e.g., in Ref.~\cite{Sato:2013}.
In all simulations, spatial functions are discretized on
equidistant grid points with spacing $\Delta z = 0.4$ within a simulation
box $-1000 < z < 1000$. 
An absorbing boundary is implemented by a mask function of $\cos^{1/4}$
shape at 10\% side edges of the box. 
Each EOM is solved by the fourth-order Runge-Kutta
method with a fixed time step size (1/10000 of an optical cycle).
The kinetic energy operator is evaluated by the eighth-order
finite difference, and spatial integrations are performed by the
trapezoidal rule.
The initial (ground state) wavefunction is obtained by the imaginary time
propagation of EOMs for each method.

\subsection{Comparison of TD-RHF and TD-EHF methods
\label{subsec:numerical_comparison}}

First we assess the performance of TD-RHF and TD-EHF methods by using
numerically exact TDSE simulation as a reference.
By doing so, we discuss the physical origin behind the well-known
failure of the TD-RHF method in describing the TI.
We used a six-cycle laser pulse of wavelength 780 nm and
intensity 8$\times$10$^{14}$ W/cm$^2$ with a trapezoidal envelope
(turning on and off linearly within two cycles). This is the same laser
profile as used in Ref.~\cite{Dahlen:2001}.

\begin{figure}[!b]
\centering
\includegraphics[width=1.0\linewidth]{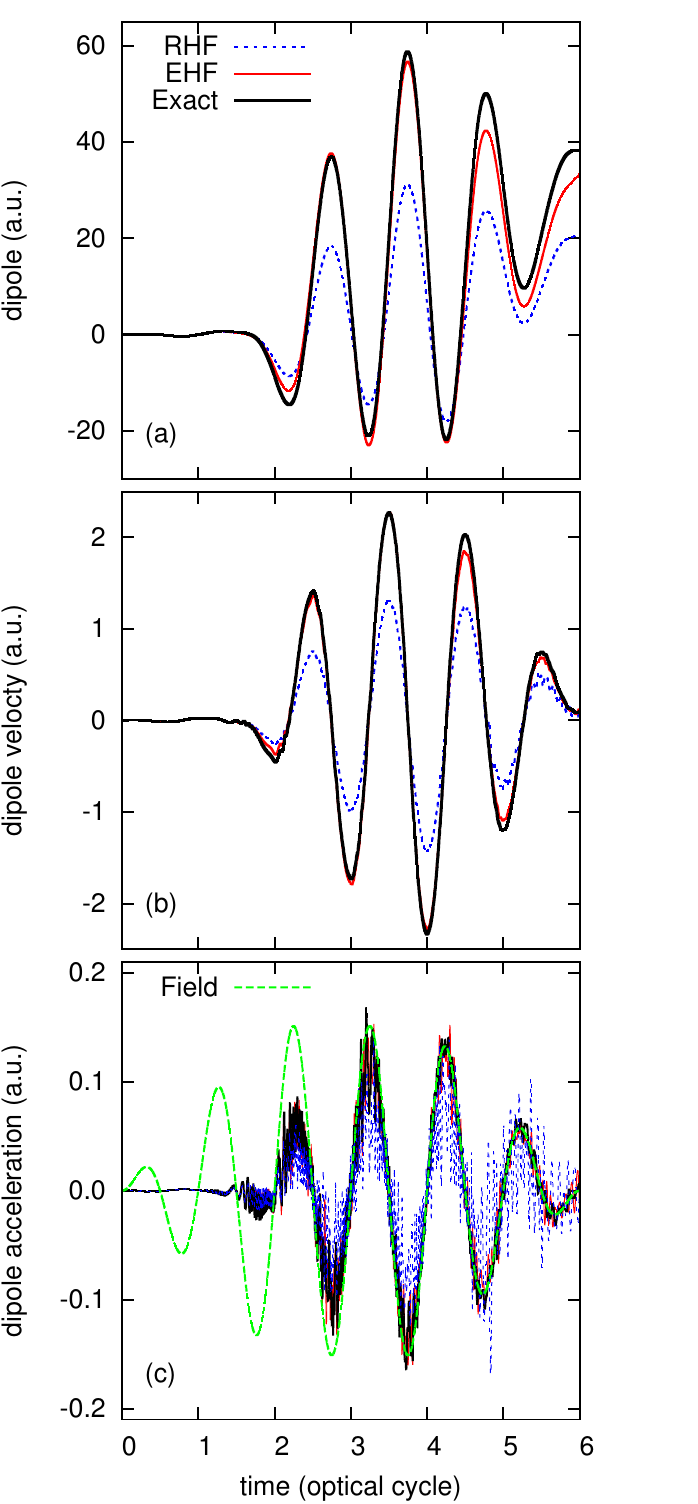}
\caption{\label{fig:prop}
Time evolution of the (a) dipole, (b) velocity, and (c) acceleration of
electrons computed with TD-RHF, TD-EHF, and TDSE methods. The laser
electric field $E(t)$ is also shown to be compared with the oscillation
of acceleration.}%
\end{figure}

In Fig.~\ref{fig:prop} we show the evolution of the expectation
values of the dipole moment (a), the velocity (b), and the acceleration
(c) of electrons defined as follows: 
\begin{subequations}\label{eq:prop}
\begin{eqnarray}
\label{eq:dipole}
\langle z \rangle &=&
\langle\Psi\vert
\sum_{i=1}^2 z_i
\vert\Psi\rangle, \\
\label{eq:velocity}
\frac{d\langle z \rangle}{dt} &=& -{\rm  i}
\langle\Psi\vert
\sum_{i=1}^2\frac{\partial}{\partial z_i}
\vert\Psi\rangle, \\
\label{eq:acc}
\frac{d^2\langle z \rangle}{dt^2} &=&
\langle\Psi\vert
\sum_{i=1}^2 
\left[\frac{z_i}{\|z_i\|^2} + E(t) \right]
\vert\Psi\rangle,
\end{eqnarray}
\end{subequations}
where the Ehrenfest expressions are used for 
the velocity and the acceleration.

As seen in the figure, the TD-RHF method underestimates
the oscillation amplitude of both dipole and dipole velocity
compared to those of TDSE. 
For the dipole acceleration, the TD-RHF method underestimates
the amplitude of the oscillation synchronized to the laser
electric field, while exaggerates the higher order (higher frequency)
response at the later stage of the pulse. 
In contrast, the TD-EHF results show reasonable agreement with TDSE
ones for these quantities.

To clarify the qualitative difference of TD-RHF and TD-EHF 
results in Fig.~\ref{fig:prop}, 
we perform the orbital decomposition analysis of the one electron probability density
$\rho(z) \equiv 2\int dz^\prime
\vert\Psi(z,z^\prime)\vert^2$ for the EHF wavefunction:
\begin{eqnarray}\label{eq:rho_ehf} 
\rho(z) = \rho_1(z) + \rho_2(z) + \rho_{12}(z),
\end{eqnarray}
where $\rho_1 \equiv \vert\psi_1\vert^2 / (1 + \vert\lambda\vert^2)$,
$\rho_2 \equiv \vert\psi_2\vert^2 / (1 + \vert\lambda\vert^2)$, and
$\rho_{12} \equiv 2 {\textrm Re} \left[\psi_1(z)\psi^*_2(z)
\lambda \right] / (1 + \vert\lambda\vert^2)$.
Accordingly, the expectation value of any local one-electron operator
$h$ is decomposed into contributions from each orbital
$\langle h \rangle_1 = {\rm tr} \left[h\rho_1\right]$,
$\langle h \rangle_2 = {\rm tr} \left[h\rho_2\right]$, and
cross term $\langle h \rangle_{12} = {\rm tr} \left[h\rho_{12}\right]$.

\begin{figure}[!t]
\centering
\includegraphics[width=1.0\linewidth]{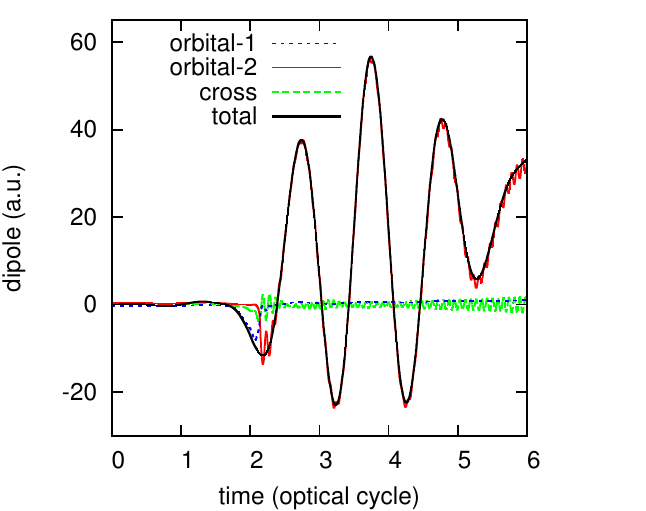}
\caption{\label{fig:dip_ehf}Orbital decomposition of the dipole moment
 of TD-EHF method. Shown are the contributions from each orbital
 $\langle z \rangle_1$ and $\langle z \rangle_2$  (orbital-1 and
 orbital-2, respectively), the cross term $\langle z \rangle_{12}$
 (cross) and the total sum, as a function of time. See text for more
 details.}%
\end{figure}

Figure~\ref{fig:dip_ehf} shows such a decomposition of the
TD-EHF dipole moment, $\langle z \rangle = \langle z \rangle_1 + \langle
z \rangle_2 + \langle z \rangle_{12}$.
One immediately finds that the large amplitude motion at $t > 2.5T$ 
is dominated by the $\psi_2$'s contribution, while the orbital
$\psi_1$ remains localized.
Since two orbitals overlap only weakly in this time region, $\langle z
\rangle_1$ and $\langle z \rangle_2$ can be interpreted as mean
positions of an electron in $\psi_1$ and $\psi_2$, respectively,
i.e., the core electron and tunnel-ionizing electron.
In this situation, 
where $\lambda \equiv \langle\psi_1|\psi_2\rangle \approx 0$, and
the effect of exchange potential $K$ is negligible in
Eqs.~(\ref{eq:td-ehf})-(\ref{eq:ehf_meanfield}),
the ionizing electron in the outer orbital $\psi_2$ feels the nuclear
potential screened by the core electron in the inner orbital $\psi_1$;
\begin{equation}\label{eq:pot_screened} 
V_{\rm eff}(z) = -\frac{2}{\|z\|} + J[\psi_1(t)](z),
\end{equation}
which, due to the localization of the inner orbital
(Fig.~\ref{fig:dip_ehf}), asymptotically approaches to the
cationic potential $V_{\rm eff}(z) \rightarrow -1/\vert z \vert$ in the
single ionization limit. 

This is completely different from the TD-RHF picture, which forces two
electrons to travel on the same orbital evolving in the field of
nuclear potential screened only inefficiently by the self Coulomb
potential [Eq.~(\ref{eq:td-rhf})], thus fails to distinguish the
core and ionizing electrons.
Effectively this causes the decreased (increased) probability density
$\rho(z)$ at large (small) $\vert z \vert$ regions compared to the exact
density, leading the underestimation of the dipole and dipole velocity
[Fig.~\ref{fig:prop} (a,b)] which is dominated by the motion of the
ionizing electron at large $\vert z \vert$, as well as the enhanced
higher order response in the dipole acceleration [Fig.~\ref{fig:prop}
(c)] to which the core electron at small $\vert z \vert$ contributes
strongly.

\subsection{Interpretation of EHF orbitals in tunneling ionization
\label{subsec:numerical_interpretation}}

\begin{figure}[!b]
\centering
\includegraphics[width=1.0\linewidth]{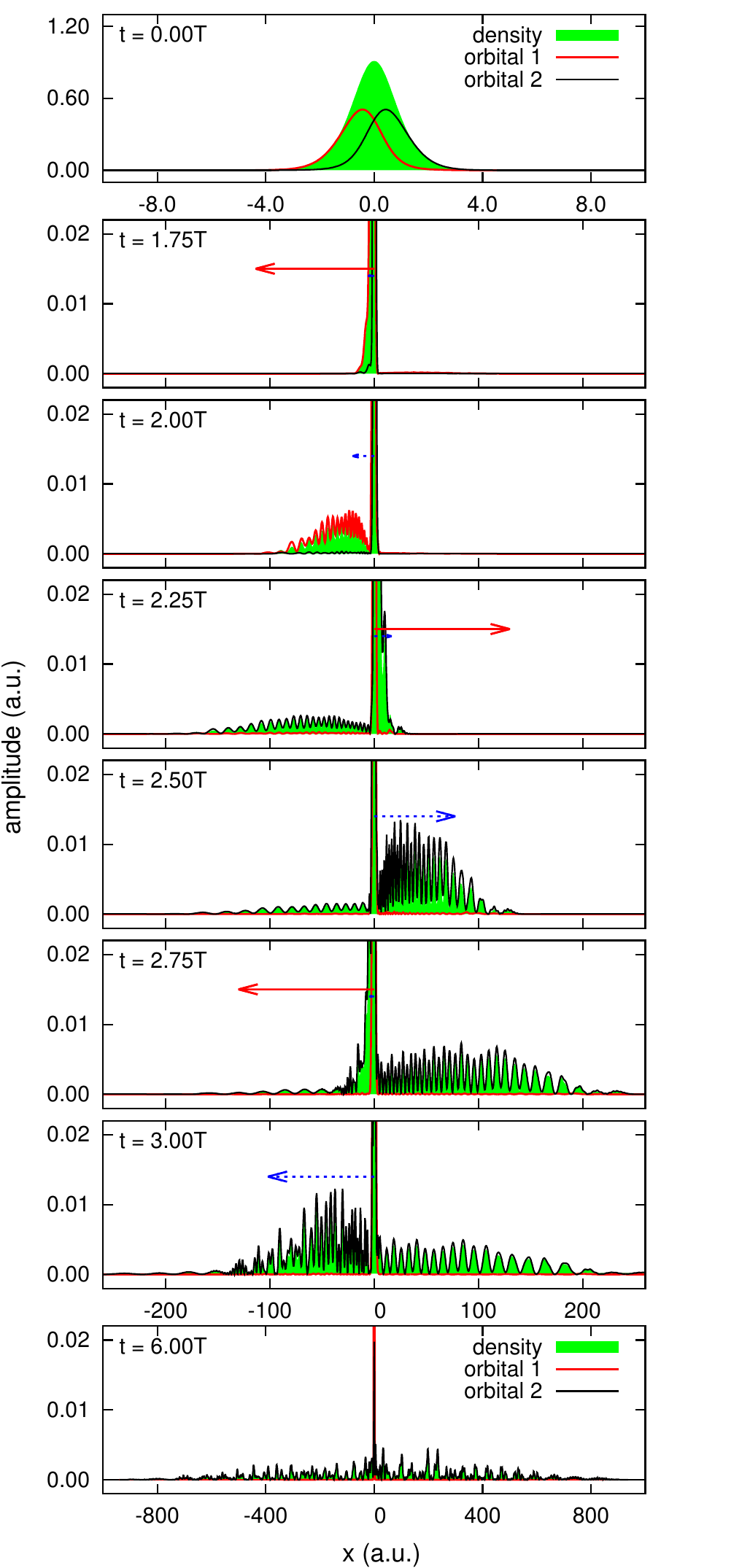}
\caption{\label{fig:orb}Snapshots of evolving EHF orbitals at
 several instances of time. Note the different length scales for the
 initial (top), the final (bottom), and the intermediate
 (middles) times. The one  electron probability density is also given by the filled region.
 The solid (red) and break (blue) arrows indicate the field force
 $-E(t)$ and the expectation value of the velocity $\langle \dot{z}
 \rangle$, respectively, in an arbitrary unit.}%
\end{figure}

\begin{figure}[!b]
\centering
\includegraphics[width=1.0\linewidth]{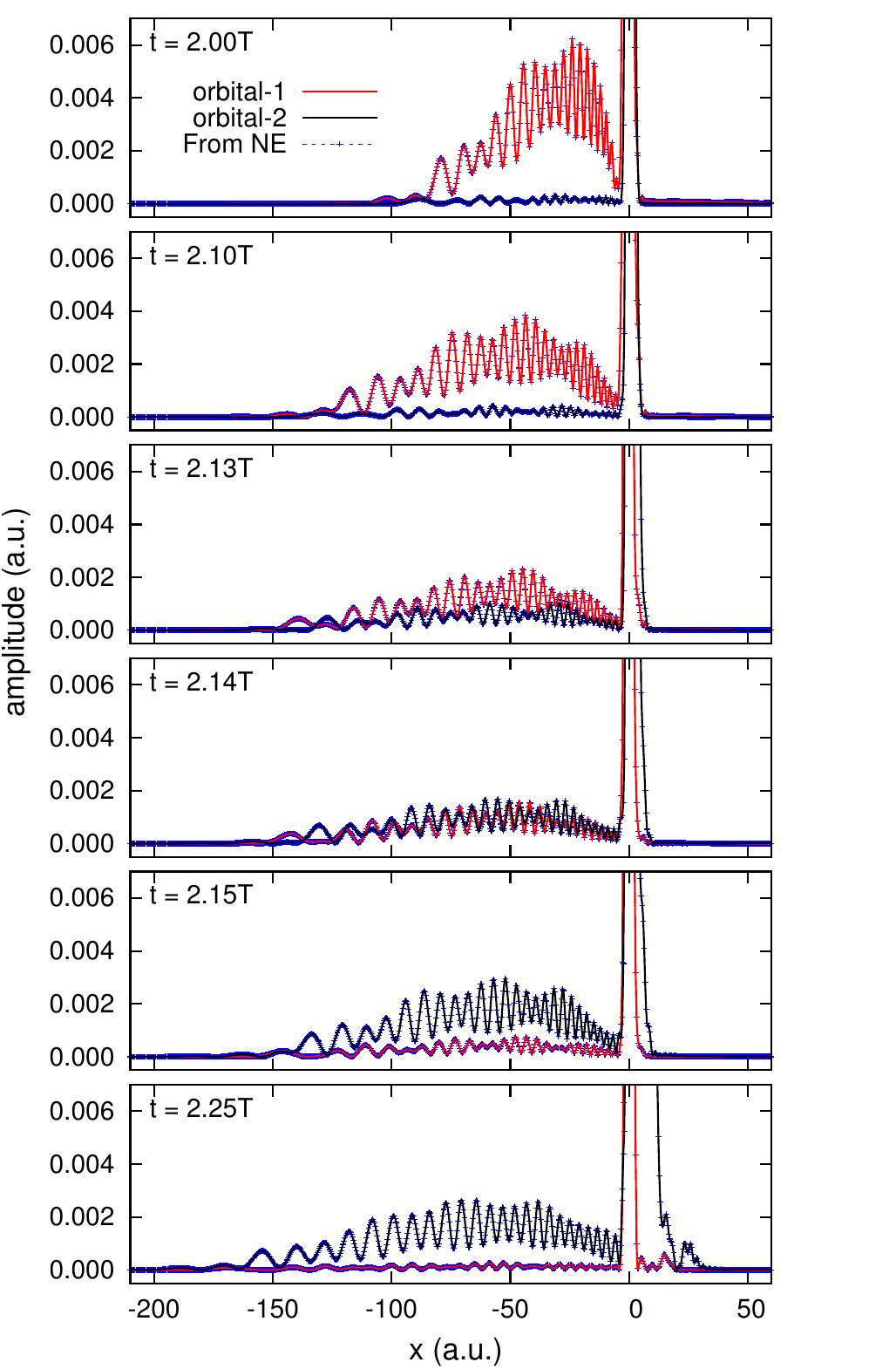}
\caption{\label{fig:orb_fine}Snapshots of evolving EHF orbitals at
 several moments of time within $2T \leq t \leq 2.25T$. The solid lines
 show two nonorthogonal orbitals directly propagated by the TD-EHF
 equation, Eq.~(\ref{eq:td-ehf}). The dashed lines show
 nonorthogonal orbitals obtained by the transformation of
 Eq.~(\ref{eq:ehf_transb}), using NCs and NOs propagated by TD-NO2
 equations, Eqs.~(\ref{eq:td-no_nc}) and (\ref{eq:td-no_no}).}%
\end{figure}

Next in Fig.~\ref{fig:orb}, we look into the time evolution of the
non-orthogonal EHF orbitals.
The initial orbitals [top panel of Fig.~\ref{fig:orb}], 
start to oscillate following the laser field. 
Until around $t < 1.5T$, the oscillation is small in magnitude
(relative to the scale of Fig.~\ref{fig:orb} at $t > 0$), 
and leads to only marginal ionization, keeping the
relative ``left'' (orbital-1) and ``right'' (orbital-2) 
location of two orbitals.
At $t = 1.75T$ when the laser force points to the negative direction of
$z$-axis with $\sim90\%$ of the maximum intensity, the ``left'' tail of
the orbital-1 deforms appreciably, leading the first
significant tunneling ionization at around $t = 2T$ when the laser force
vanishes.
Curiously at the first look, 
the outgoing part of orbital-1 (red) at $t = 2T$ looks replaced
by the delocalized orbital-2 (black) at $t = 2.25T$.
A closer inspection of the orbital evolution within this time region
(Fig.~\ref{fig:orb_fine}) shows that there occurs a strong mixing and
character change between two orbitals. This effect is also visible in
the orbital components of the dipole in Fig.~\ref{fig:dip_ehf}.

After the character mixing, the two orbitals get
clearly characterized as ``inner'' (red) and ``outer'' (block)
orbitals, the former being localized at origin, while the latter being
delocalized.
At $t = 2.25T$, the ``right'' tail of the outer orbital
deforms toward positive direction, while the left-going part of the
orbital loses stream feeling the reverse acceleration. 
At another 1/4 cycle later ($t = 2.5T$) when the laser force vanishes
again, the whole wave packet drives with a positive velocity, undergoing
both tunneling ionization towards ``right'' and recombination or
recollision at the origin from the ``left''. 
At $t = 2.75T$, one sees a mirrored situation
of that at half cycle before, namely, deformation of the ``left'' tail
and nearly standing ``right'' wave packet. 
Then at $ t = 3T$ the outer orbital represents both tunneling ionization
towards ``left'' and recombination/recollison from the ``right''.
During the latter half of the pulse $(3T < t < 6T)$, the ionizing
electron in the outer orbital evolves like a single active electron
feeling the screened nuclear potential $V_{\rm eff}$ of
Eq.~(\ref{eq:pot_screened}), resulting in the wide spread continuum
state at the end of the pulse [bottom of Fig.~(\ref{fig:orb})].

These analyses of EHF orbitals demonstrate the
interpretive power of the TD-EHF method, which provides intuitive
picture for the TI event and subsequent electron dynamics compatible to
the semi-classical three step model \cite{Corkum:1993}. 
Now we point out that the same picture can be drawn from computationally
more convenient TD-NO2 method as shown in Fig.~\ref{fig:orb_fine}.
Here we solve Eq.~(\ref{eq:td-no}) for NCs $\{A_1, A_2\}$ and NOs
$\{\phi_1,\phi_2\}$, from which the non-orthogonal orbitals
$\{\psi_1,\psi_2\}$ are computed at each time step using
Eq.~(\ref{eq:ehf_transb}). The obtained instantaneous non-orthogonal
orbitals, plotted with dashed lines in Fig.~\ref{fig:orb_fine},
perfectly match the EHF orbitals. This numerically confirms the
equivalence of the two sets of equations Eq.~(\ref{eq:td-ehf}) and
Eq.~(\ref{eq:td-no}). 
The TD-NO2 method has some practical advantages over the TD-EHF method,
e.g., allowing larger time steps and smaller simulation box sizes for
the numerically convergent simulation.
The direct propagation of nonorthogonal orbitals in the TD-EHF method,
especially with unnormalized orbitals \cite{Dahlen:2001, Nguyen:2006}, 
seems to occasionally cause numerical difficulties
as shown in Appendix~\ref{app:ehf_vs_no2}.

\subsection{High harmonic generation spectrum
\label{subsec:numerical_hhg}}

\begin{figure}[!b]
\centering
\includegraphics[width=1.0\linewidth]{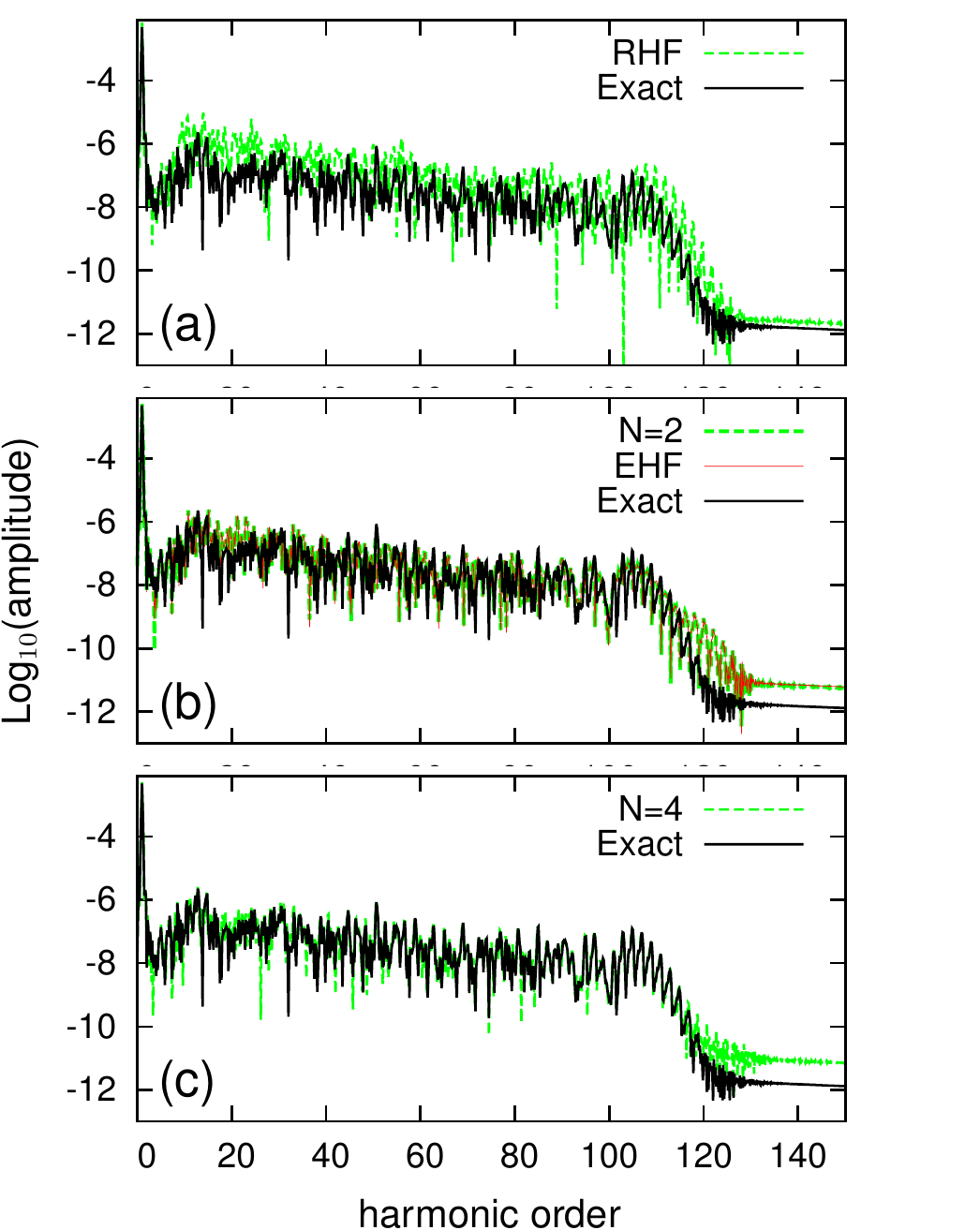}
\caption{\label{fig:hhg}HHG spectra obtained by TD-RHF (a),
TD-EHF/TD-NO2 (b), and TD-NO4 (c) methods. The TDSE spectrum
is shown in each panel for comparison.}%
\end{figure}

Figure~\ref{fig:hhg} compares the HHG spectra computed with TD-RHF,
TD-NO2, TD-NO4, and TDSE methods.
The spectrum is defined as the Fourier
transform of the dipole acceleration in Eq.~(\ref{eq:acc}) neglecting
the bare field term.
From a three-step model analysis \cite{Corkum:1993} with Hartree-Fock-Koopmans
ionization potential, the cutoff is predicted at about 103rd harmonic,
in a good agreement with the TDSE spectrum.

As seen in the figure, the TD-RHF method overestimates the amplitude
except for a first few harmonics. This reflects the enhanced high order
responses in the dipole acceleration in Fig.~\ref{fig:prop}~(c), caused
by unphysical closed-shell description of two electrons.
The panel (b) shows the perfect agreement between TD-EHF and TD-NO2
results, again confirming the equivalence of these theories.
The TD-NO2 (TD-EHF) spectrum shows a better agreement with the TDSE
one, than the TD-RHF spectrum, both in the amplitude on the plateau and in the cutoff position.
However, it fails to reproduce finer structures in the spectrum,
especially at and beyond the cutoff frequency.
It is always difficult to make a confident prediction of a given
observable with a single approximate method.
In this respect, the possibility of the systematic improvement in the
TD-NO$n$ series of approaches is highly appreciated, allowing a
convergence study and self checking on the validity of the
approximation. The TD-NO4 spectrum in Fig.~\ref{fig:hhg}~(c) shows 
a quite good agreement with the exact one.

\subsection{Nonsequential double ionization
\label{subsec:numerical_nsdi}}

The HHG spectrum, and all the one particle properties investigated thus far,
are not very sensitive to the individual motion of electrons since
they depend only on the average motion of two electrons. 
To make a severer assessment on the quality of the two electron
wavefunction, we investigate the NSDI process, which crucially
depends on the motion of each electron.
The TD-EHF method has been applied to the NSDI of 1D-He
\cite{Dahlen:2001} and 1D-hydrogen molecule \cite{Nguyen:2006} models.
Here we extend these previous works to take deeper account of roles of
the electron correlation.
We used a three cycle pulse with wavelength 750 nm, $\sin^2$
envelope, and intensities varied from 2$\times$10$^{14}$ to
3$\times$10$^{15}$ W/cm$^2$. 
Following Refs.~\cite{Dahlen:2001, Nguyen:2006}, the ionization yields
are estimated from the two electron probability $\rho(z_1,z_2) =
\Psi(z_1,z_2)\Psi^*(z_1,z_2)$ as
\begin{subequations}\label{eq:ion}
\begin{eqnarray}
P_0 &=&
\int_{|z_1|<R}\hspace{-1em}dz_1\int_{|z_2|<R}\hspace{-1em}dz_2 \rho(z_1,z_2), \\
P_1 &=& 2
\int_{|z_1|>R}\hspace{-1em}dz_1\int_{|z_2|<R}\hspace{-1em}dz_2 \rho(z_1,z_2), \\
P_2 &=&
\int_{|z_1|>R}\hspace{-1em}dz_1\int_{|z_2|>R}\hspace{-1em}dz_2 \rho(z_1,z_2),
\end{eqnarray}
\end{subequations}
where $P_0$, $P_1$, and $P_2$ are interpreted as zero (no ionization),
single, and double ionization probabilities, respectively, with $R = 18$.

\begin{figure}[!t]
\centering
\includegraphics[width=1.0\linewidth]{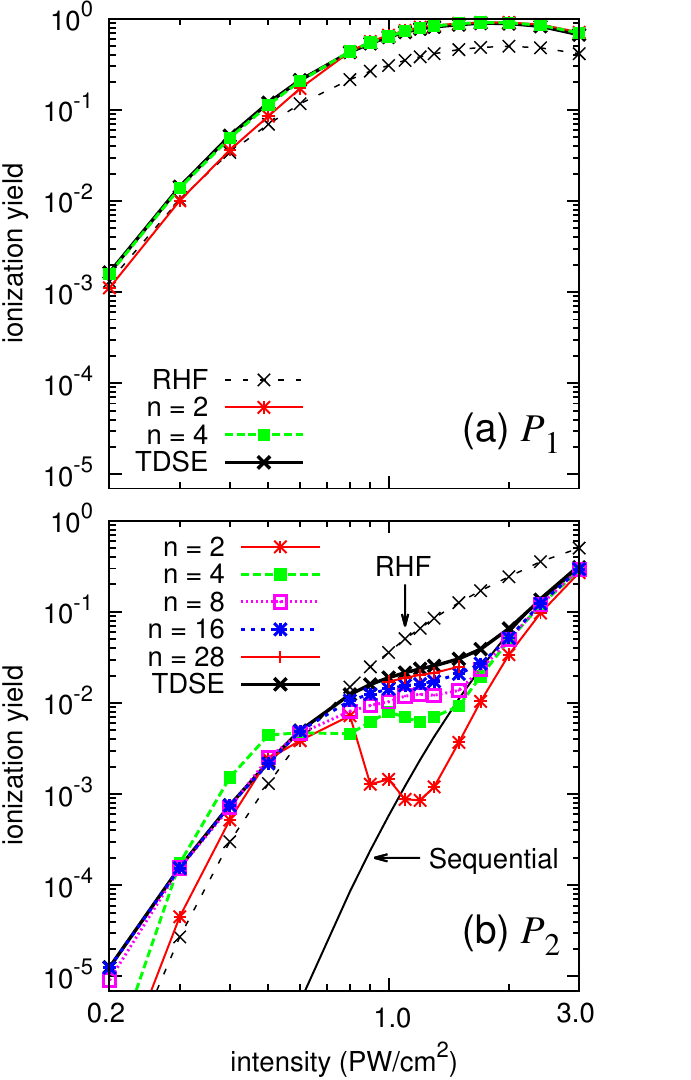}
\caption{\label{fig:nsdi_ion}Single and double ionizations of
 1D-He obtained by various methods. Single (a) and (b)
 double ionization probabilities as a function of peak intensity. Also
 shown in (b) is the ionization probability of He$^+$, corresponding to the
 sequential model prediction of the double ionization probability.}%
\end{figure}

Figure~\ref{fig:nsdi_ion} shows the intensity dependence of $P_1$
and $P_2$ at the end of the pulse.
In accordance with previous 1D simulations
with longer pulses \cite{Dahlen:2001, Nguyen:2006}, 
the present simulation properly reproduces the experimental features;
the exact (TDSE) result shows the ``shoulder'' like enhancement 
relative to the sequential result at the intensity region
$0.8\times10^{14} \sim 1.5\times10^{15}$ W/cm$^2$ (shoulder region).
The TD-RHF gives radically different results for both $P_1$ and
$P_2$ compared to the TDSE results, with $P_1$ strongly underestimated
while $P_2$ unphysically overestimated at and above the shoulder region.
Both of these wrong behaviors originate from the closed-shell
ansatz of Eq.~(\ref{eq:rhf}), which in the first place cannot distinguish
sequential and nonsequential processes.
%
The TD-NO$n$ series show a rapid convergence for $P_1$ as is the case for the 
one-particle observables in preceding sections.
The TD-NO4 method already gives a nearly convergent curve of $P_1$
against the intensity, with relative deviations from the TDSE values
less than 5.5\%. 

In a striking contrast, Fig.~\ref{fig:nsdi_ion}~(b) reveals the
extremely slow convergence of $P_2$ with respect to the number of NOs.
Although the $P_2$ curve of the TD-NO2 method at least reproduces
the shoulder feature, 
the $P_2$ values at around $10^{15}$ W/cm$^2$ are underestimated by an
order of magnitude.
A moderate overall accuracy is first achieved by TD-NO8, albeit with a
sizable error at the shoulder region. The largest relative deviations of
TD-NO$n$ values of $P_2$ from the TDSE one within the shoulder region
are 74\%, 55\%, 32\%, and 18\% with $n =$ 4, 8, 16, and 28,
respectively.
Such slow convergence is a consequence of highly correlated electron
dynamics involved in the NSDI; the recollision between the returning
electron and parent ion requires precise account of instantaneous,
short-range electron-electron correlation.

The recollision induced NSDI process can be separated into 
(i) the direct recollisional ionization, dominant at the shoulder region and 
(ii) recollision-excitation and subsequent tunneling ionization,
dominant at lower intensities.
In the first mechanism, the high kinetic energy of the returning
electron permits a closer approach
[than in the mechanism (ii)]
of electrons surpassing the electron-electron repulsive potential, and 
the direct energy transfer through the potential.
This causes the especially slow convergence at the shoulder region.
On the other hand, the lower energy inelastic collision in the second mechanism
prevents two electrons from coming too close [in contrast to the
mechanism (i)]. This is reflected by a faster convergence of $P_2$
values at lower intensities in Fig.~\ref{fig:nsdi_ion}~(b).
The convergence becomes faster again at the highest intensities, where
the electron-electron interaction is less important relative to the
strong electron-laser interaction.

\subsection{Roles of electron correlation in TI, HHG, and NSDI
\label{subsec:numerical_correlation}}

\begin{figure}[!b]
\centering
\includegraphics[width=1.0\linewidth]{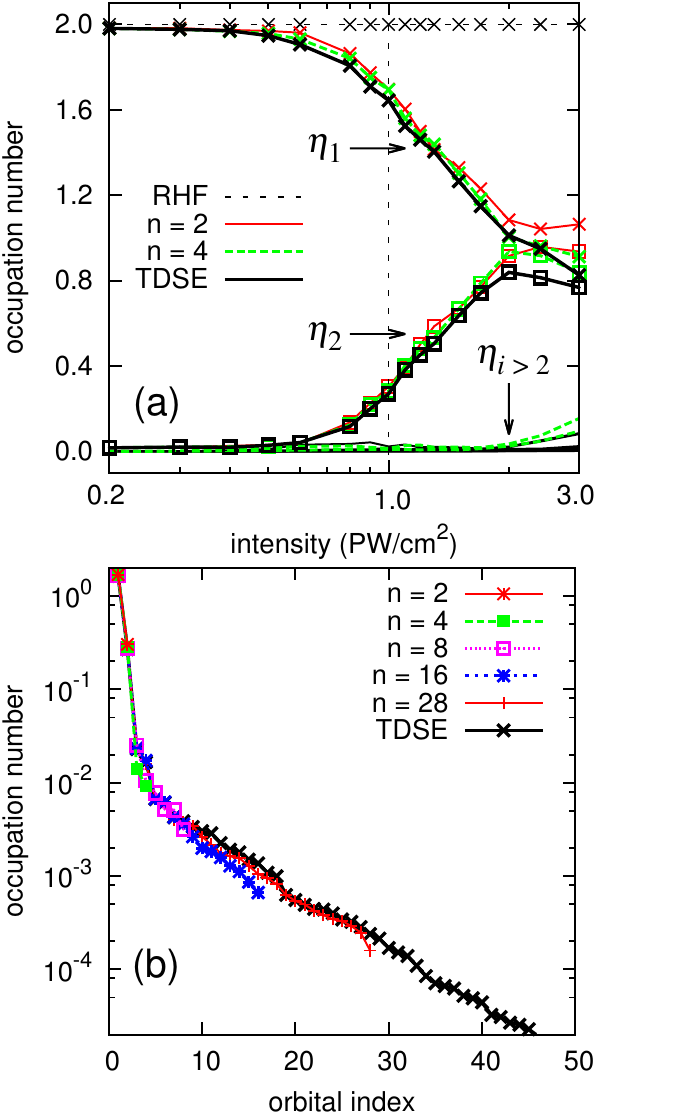}
\caption{\label{fig:nsdi_occ}(a) Natural occupation numbers (ONs)
 at the end of the pulse as a function of peak  intensity. Shown are the
 available number of ONs; one for TD-RHF which is just a constant $2.0$,
 $n$ for TD-NO$n$, and 5000 (number of grid points) for TDSE
 methods. 
The first and second ONs are labeled with cross ($\times$) and
square ($\square$) markers, respectively.
(b) ONs plotted in the descending order for TD-NO$n$ and TDSE
methods at the end of the pulse of intensity 1 PW/cm$^2$ [indicated
with the vertical line in (a)].}%
\end{figure}

Finally in this section, we summarize different roles of the electron
correlation in TI, HHG, and NSDI processes.
Figure~\ref{fig:nsdi_occ}~(a) shows the intensity dependence of the
natural occupation numbers (ONs). The ONs $\left\{\eta_i\right\}$ are
calculated at the end of the pulse of Sec.~\ref{subsec:numerical_nsdi},
by Eq.~(\ref{eq:occupation}) for TD-NO$n$, or by diagonalizing the grid
representation of the 1RDM [Eq.~(\ref{eq:1rdm_def})] for TDSE.
As clearly seen in the figure except for TDHF, the first two ONs experience a
drastic change, with decreasing (increasing) population for the
initially strongly (weakly) populated first (second) natural orbital. 

In TD-NO2, the weak ($\eta_1 \gg \eta_2$) and high
($\eta_1 \simeq \eta_2$) intensity limits of two ONs correspond,
respectively, to the strongly overlapping
($\vert\langle\psi_1\vert\psi_2\rangle\vert \simeq 1$) and nearly
orthogonal ($\vert\langle\psi_1\vert\psi_2\rangle\vert \simeq 0$) EHF 
orbitals as followed from Eq.~(\ref{eq:ehf_transb}).
Physically, this represents the transition, during the course of increasing
intensity, from the limit of no ionization to the other limit of
completed single ionization. The same transition in the time domain is nothing 
but the TI process (in a sufficiently strong field), and 
the TDHF method cannot describe this transition. 
Thus, the electron correlation (effects beyond the single determinant
description) is first manifested in the TI process itself;
the breakdown of initial closed-shell symmetry.

Figure~\ref{fig:nsdi_occ}~(b) shows the final ONs for the
field intensity $10^{15}$ W/cm$^2$, plotted in logarithmic scale in the
descending order. The ON spectrum obviously changes the slope in the
vicinity of $\eta_4$,
suggesting that a first few and remaining NOs play different roles.
To understand this behavior, we directly see errors of
one- and two-electron probability densities (hereafter called 1-density
and 2-density, respectively) of TD-NO$n$ compared to exact TDSE densities;
\begin{subequations}
\begin{eqnarray} 
\delta\rho_n(z)&\equiv&\rho_n(z)-\rho_\infty(z), \\
\delta\rho_n(z_1,z_2)&\equiv&\rho_n(z_1,z_2)-\rho_\infty(z_1,z_2),
\end{eqnarray}
\end{subequations}
where $\rho_n$ ($\rho_\infty$) is the density from TD-NO$n$ (TDSE)
wavefunction. The first and second entries of table~\ref{tab:devden}
show the absolute error $|\delta\rho_n(z)|$ of 1-density
integrated and normalized over the inner ($|z|<R$) and outer ($|z|>R$)
spatial regions; 
\begin{subequations}\label{eq:den1dev}
\begin{eqnarray}
\Delta^{(1)}_< &=& \frac{1}{N_<}
\int_{|z|<R}\hspace{-1em}dz | \delta\rho_n(z) |, 
\\
\Delta^{(1)}_> &=& \frac{1}{N_>}
\int_{|z|>R}\hspace{-1em}dz | \delta\rho_n(z) |, 
\end{eqnarray}
\end{subequations}
where $N_<$ and $N_>$ are the norms of the exact 1-density $\rho_\infty(z)$
within inner and outer regions, respectively. Table~\ref{tab:devden} shows a
rapid convergence of the 1-density, especially for the inner
region. Therefore, we conclude that the first several NOs in 
Fig.~\ref{fig:nsdi_occ}~(b), before the slope change, are responsible
for quantitative improvement of single electron dynamics (governed by
1-density) on top of the basic two orbital description of TD-NO2 
(or equivalently TD-EHF).
As a consequence, one electron properties, including HHG spectrum, are
described well by relatively small number of NOs.
\begin{table}[!t]
\caption{\label{tab:devden}
The domain divided relative errors (\%) of 1-density ($\Delta^{(1)}$) and
2-density ($\Delta^{(2)}$) relative to the TDSE densities, at the end of
the pulse with intensity $10^{15}$ W/cm$^2$.
See text and Eqs.~(\ref{eq:den1dev}) and (\ref{eq:den2dev}) for
definition of $\Delta^{(1)}$ and $\Delta^{(2)}$. The inner ($N_<$) and
outer ($N_>$) norms of the 1-density, and no ($P_0$), single ($P_1$),
and double ($P_2$) ionization parts of 2-density are also shown for TDSE.}
\begin{ruledtabular}
\begin{tabular}{cddddd}
\multicolumn{1}{r}{$n$} &
\multicolumn{1}{c}{$\Delta^{(1)}_{<}$} &
\multicolumn{1}{c}{$\Delta^{(1)}_{>}$} &
\multicolumn{1}{c}{$\Delta^{(2)}_{<<}$} &
\multicolumn{1}{c}{$\Delta^{(2)}_{><}$} &
\multicolumn{1}{c}{$\Delta^{(2)}_{>>}$} \\
\hline
\multicolumn{1}{r}{RHF: 1} & 25.6 & 88.9 & 102.0 & 91.1 & 264.2 \\
\multicolumn{1}{r}{EHF: 2} & 11.6 & 23.9 &  35.3 & 28.5 &  96.8 \\
\multicolumn{1}{r}{ 4}     &  5.4 & 10.8 &  16.0 & 17.7 &  94.8 \\
\multicolumn{1}{r}{ 8}     &  1.9 &  4.1 &   6.7 &  9.2 &  81.6 \\
\multicolumn{1}{r}{12}     &  1.2 &  3.0 &   4.5 &  6.3 &  73.9 \\
\multicolumn{1}{r}{16}     &  0.9 &  2.4 &   3.6 &  5.0 &  62.9 \\
\multicolumn{1}{r}{20}     &  0.7 &  1.9 &   2.1 &  4.3 &  55.2 \\
\multicolumn{1}{r}{28}     &  0.3 &  1.1 &   1.6 &  2.6 &  36.9 \\
 &
\multicolumn{1}{c}{$N_{<}$} &
\multicolumn{1}{c}{$N_{>}$} &
\multicolumn{1}{c}{$P_0$} &
\multicolumn{1}{c}{$P_1$} &
\multicolumn{1}{c}{$P_2$} \\
\hline
\multicolumn{1}{r}{TDSE} & 1.332 & 0.668 & 0.351 & 0.630 & 0.019 \\
\end{tabular}
\end{ruledtabular}

\end{table}

Table~\ref{tab:devden} also lists domain divided relative errors of 2-density; 
\begin{subequations}\label{eq:den2dev}
\begin{eqnarray}
\Delta^{(2)}_{<<} &=& \frac{1}{P_0}
\int_{|z_1|<R}\hspace{-1em}dz_1\int_{|z_2|<R}\hspace{-1em}dz_2 |\delta\rho(z_1,z_2)|, \\
\Delta^{(2)}_{><} &=& \frac{2}{P_1}
\int_{|z_1|>R}\hspace{-1em}dz_1\int_{|z_2|<R}\hspace{-1em}dz_2 |\delta\rho(z_1,z_2)|, \\
\Delta^{(2)}_{>>} &=& \frac{1}{P_2}
\int_{|z_1|>R}\hspace{-1em}dz_1\int_{|z_2|>R}\hspace{-1em}dz_2 |\delta\rho(z_1,z_2)|,
\end{eqnarray}
\end{subequations}
where $P_i$ $(i = 0,1,2)$ are the exact
(TDSE) probabilities from Eq.~(\ref{eq:ion}).  
The quantities $\Delta^{(2)}_{<<}$, $\Delta^{(2)}_{><}$, and
$\Delta^{(2)}_{>>}$ measure the accuracy of 2-density for electron 
pairs undergoing no, single, and double
ionization. Table~\ref{tab:devden} clearly demonstrates slower 
convergence of 2-density than 1-density, especially for the doubly ionized
pair of electrons. This reflects the strong short-range correlation in
NSDI process as discussed in Sec.~\ref{subsec:numerical_nsdi}. 
Therefore, the totality of weakly populated NOs in
Fig.~\ref{fig:nsdi_occ}~(b), after the slope change,
contributes to the improvement in the
description of the explicitly correlated relative motion of two
electrons governed by 2-density.


\section{Summary\label{sec:summary}}

This work addresses the structure of approximate two
electron wavefunction required for properly describing intense laser
driven ionization dynamics. We theoretically establish the equivalence
of TD-EHF method and MCTDHF method with two occupied orbitals. The
latter method is formulated in the form of natural expansion, allowing
the direct propagation of NOs. The time-dependent NOs can be
transformed back into the nonorthogonal EHF orbitals, thus combines the
computational advantages of orthogonal orbitals and interpretive power
of nonorthogonal orbitals.

Numerical application of TD-NO$n$ approaches to 1D-He model highlights
different manifestations of electron correlation.
(i) The breaking closed-shell symmetry during the TI
process, of which inclusion is essential (by TD-EHF or TD-NO2 as the simplest
approximation) for a physically meaningful description. 
(ii) The correlation correction to the one electron density, 
(included by a few more orbitals) which quantitatively refines the
description of single electron processes (in the presence of other
electrons), thus important for accurate computations of one electron
properties such as HHG.  
(iii) The instantaneous, short-range electron-electron correlation 
(requiring much more orbitals) involved in the recollision-induced NSDI process. 
The NO-based reformulation of the TD-EHF method enables such 
systematic evaluations of roles of electron correlation in TI, HHG, and
NSDI processes, in a unified variational ansatz of Eq.~(\ref{eq:mctdhf_no}).

There exists a promising extension of the NO-based approach
to multielectron systems in the following form;
\begin{eqnarray}\label{eq:apg}
\Psi(\bm{x}_1,\bm{x}_2,\cdot\cdot\cdot,\bm{x}_N) = \hat{\mathcal{A}} 
\prod_{P=1}^{N/2}
\psi_P(\bm{x}_{2P-1},\bm{x}_{2P}),
\end{eqnarray}
where the $N$-electron wavefunction (assuming even $N$ for simplicity)
is given in terms of antisymmetrized product of {\it geminals} (two
electron wavefunction), instead of orbitals as in TDHF method.
Each geminal $\psi_P$ can take the form of the natural expansion in
Eq.~(\ref{eq:mctdhf_no}). 
The geminal product wavefunction of Eq.~(\ref{eq:apg}) has a long
history in the stationary quantum chemistry, particularly with the so
called strong orthogonality condition \cite{Arai:1960, Kutzelnigg:1964},
known under the name of antisymmetrized products of strongly orthogonal
geminals \cite{Kutzelnigg:1964}, or generalized valence-bond in the
perfect-paring approximation \cite{Hunt:1972}. These methods are
computationally far less demanding than configuration-based approaches
like MCTDHF method, and at the same time, keep the conceptual simplicity
of EHF wavefunction for two electron systems. A time-dependent theory
based on Eq.~(\ref{eq:apg}) is now under development.

\begin{acknowledgments}
 This research is supported in part by Grant-in-Aid for Scientific
 Research (No. 23750007, No. 23656043, No. 23104708,
 No. 25286064, No. 26390076, and No. 26600111) from the
 Ministry of Education, Culture, Sports, Science and Technology (MEXT)
 of Japan, and also by Advanced Photon Science Alliance (APSA) project
 commissioned by MEXT.
 This research is also partially supported by the Center of Innovation Program
 from Japan Science and Technology Agency, JST.
\end{acknowledgments}

\appendix
\section{Numerical stability of TD-EHF and TD-NO$2$ methods\label{app:ehf_vs_no2}}
\begin{figure}[!b]
\centering
\includegraphics[width=1.0\linewidth]{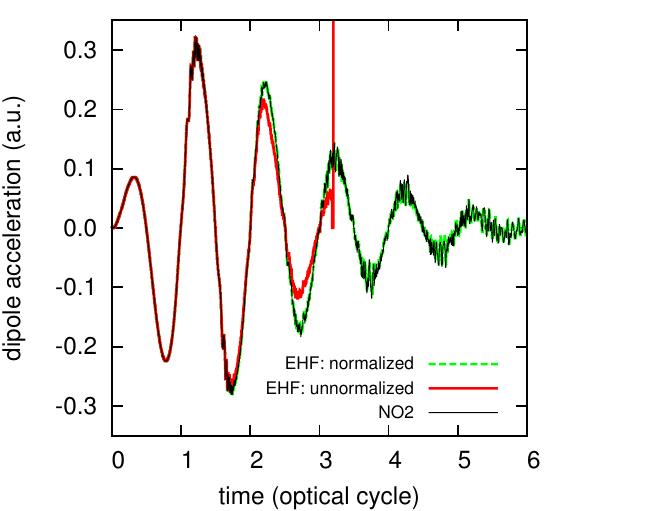}
\caption{\label{fig:acc_ehf}
Time evolution of the dipole acceleration in case of an increased peak
intensity, 3$\times$10$^{15}$ W/cm$^2$ and a reduced box size. 
Results of TD-EHF with normalized and unnormalized orbitals, and TD-NO$2$.}%
\end{figure}

As discussed in Secs.~\ref{subsec:td-ehf} and \ref{subsec:ehf_orth}, the
TD-EHF method either with normalized orbitals \cite{Pindzola:1995,
Pindzola:1997, Tolley:1999} adopted in this work or with unnormalized orbitals
\cite{Dahlen:2001, Nguyen:2006}, and the TD-NO$2$ method are
mathematically equivalent.
We found, however, that the propagation of TD-EHF equation, especially
with unnormalized orbitals, requires severer simulation condition 
(larger simulation box and/or finer temporal step size).
Figure~\ref{fig:acc_ehf} shows the dipole acceleration as a function of
time with the same simulation condition as in 
Sec.~\ref{subsec:numerical_comparison} except for the increased peak 
intensity of 3$\times$10$^{15}$ W/cm$^2$ and smaller simulation box,
$|z| < 600$. The TD-EHF simulation with unnormalized orbitals uses EOMs
given in Ref.~\cite{Nguyen:2006}.
It is confirmed that this box size is sufficient for obtaining the
convergent acceleration value with TD-NO2.
As seen in the figure the TD-EHF
result with unnormalized orbitals begins to deviate from the other results
after two optical cycles, and gets unstable after three cycles.
This behavior originates from the presence of orbital overlap integrals
in the TD-EHF equation [Eq.~(\ref{eq:td-ehf})]. 
It requires a precise conservation of orbital
norms and mutual overlap {\it within} the simulation box (therefore
necessitates a large box to support the outgoing electrons); otherwise the
inaccurate $\bm{S}$ matrix of Eq.~(\ref{eq:ehf_ovlp}) alters the
dynamics even at the core region (responsible for the dipole
acceleration). The normalized orbital approach of TD-EHF largely
alleviates this problem since it only involves overlap $\lambda \equiv
\langle\psi_1|\psi_2\rangle$ in Eq.~(\ref{eq:ehf_ovlp}) with each
orbital {\it assumed} normalized. 

In the TD-NO2 formulation, the orbital
propagation is more stable since it utilizes the {\it assumed} orbital
orthonormality, not involving box normalization in its EOM [Eq.~(\ref{eq:td-no_no})].
Instead, the normalization of the total wavefunction is
governed by the EOM for the NCs [Eq.~(\ref{eq:td-no_nc})], which is
unitary irrespective of a box truncation. As a result, the TD-NO
approach allows a relatively small simulation box (with a good
absorbing boundary) for the dynamics at the core region. It is also
compatible to the advanced treatment of outgoing flux as developed in
Ref.~\cite{Scrinzi:2012}.

\bibliography{refs.bib}
\end{document}